\documentclass{iau} 
\usepackage{graphicx}

\title[S~344.~~Neutral Hydrogen in Nearby Dwarf Galaxies]
      {Neutral Hydrogen in Nearby Dwarf Galaxies}

\author[B\"arbel S. Koribalski] 
{B\"arbel S. Koribalski}

\affiliation{CSIRO Astronomy and Space Science, 
   Australia Telescope National Facility \\ PO Box 76, Epping, NSW 1710,
   Australia \\ email: {\tt Baerbel.Koribalski@csiro.au}} 

\pubyear{2018}
\volume{344}  
\jname{Dwarf Galaxies: From the Deep Universe to the Present}
\editors{K. McQuinn \& S. Stierwalt, eds.}
\begin{document}

\maketitle

\begin{abstract}
Here I briefly highlight our studies of the gas content, kinematics and star 
formation in nearby dwarf galaxies ($D < 10$ Mpc) based on the {\bf `Local 
Volume H\,{\sc i} Survey'} (LVHIS, Koribalski et al. 2018), which was 
conducted with the Australia Telescope Compact Array (ATCA). 
The LVHIS sample consists of nearly 100 galaxies, including new discoveries, 
spanning a large diversity in size, shape, mass and degree of peculiarity. 
The hydrogen properties of dwarf galaxies in two nearby groups, Sculptor and 
CenA / M83, are analysed and compared with many rather isolated dwarf 
galaxies. Around 10\% of LVHIS galaxies are transitional or mixed-type 
dwarf galaxies (dIrr/dSph), the formation of which is explored. --- I also 
provide a brief update on {\bf WALLABY Early Science}, where we focus on 
studying the H\,{\sc i} properties of galaxies as a function of environment. 
WALLABY (Dec $< +30$ degr, $z < 0.26$) is conducted with the Australian SKA 
Pathfinder (ASKAP), a $\sim$6-km diameter array of $36 \times 12$-m dishes, 
each equipped with wide-field (30 sq degr) Chequerboard Phased Array Feeds.  

\end{abstract}

\section{LVHIS -- The Local Volume H\,{\sc i} Survey}
\vspace{0.3cm}

The LVHIS project provides high-resolution H\,{\sc i} spectral line and 
20-cm radio continuum data products for nearly 100 nearby galaxies, based 
on over 2500 hours of ATCA observations. The raw data and the data products 
are available for download. The LVHIS galaxy properties are presented by 
\cite[Koribalski et al. (2018)]{koribalski18}, together with an overview, 
H\,{\sc i} galaxy atlas and a short description of each galaxy. The 
star-formation properties of LVHIS galaxies are analysed by \cite[Wang et 
al. (2017)]{wang17}, using multi-wavelength images, and \cite[Shao et al. 
(2018)]{shao18} using the ATCA radio continuum data. The LVHIS Galaxy Atlas 
and database (incl. FITS files) can be found on-line at 
{\em www.atnf.csiro.au/research/LVHIS}; more products will be
added in future.

Within LVHIS, dwarf galaxies are in the vast majority, including transitional, 
irregular and Magellanic barred galaxies. For detailed studies of individual 
LVHIS galaxies see, for example, \cite[van Eymeren et al. (2010)]{e2010} on 
IC\,4662 and NGC~5408, \cite[L\'opez-S\'anchez et al. (2012)]{ls12} on the 
peculiar starburst dwarf galaxy NGC~5253, \cite[Westmeier et al. (2011, 
2013)]{w11,w13} on NGC~55 and NGC~300, 
\cite[For et al. (2012)]{for12} on the Circinus Galaxy, and \cite[Koribalski 
\& L\'opez-S\'anchez (2009)]{kl09} on the NGC~1512/1510 system. In the LVHIS 
overview paper (\cite[Koribalski et al. 2018]{koribalski18}) we highlight 
the nearest known neighours to each galaxy and study their 3D environment, 
made possible by independent distance estimates now available for most Local 
Volume galaxies. On several occasions we discovered uncatalogued dwarf 
companions to the LVHIS galaxies, and we expect many more discoveries of dwarf 
galaxies in future, large-scale SKA and SKA pathfinder H\,{\sc i} survey such 
as WALLABY (\cite[Koribalski 2012]{koribalski12}; see Section~2), the ASKAP 
H\,{\sc i} All-Sky Survey (Dec $< +30$ degr; $z < 0.26$).  \\

\begin{figure}[h] 
\begin{center}
\includegraphics[width=14cm]{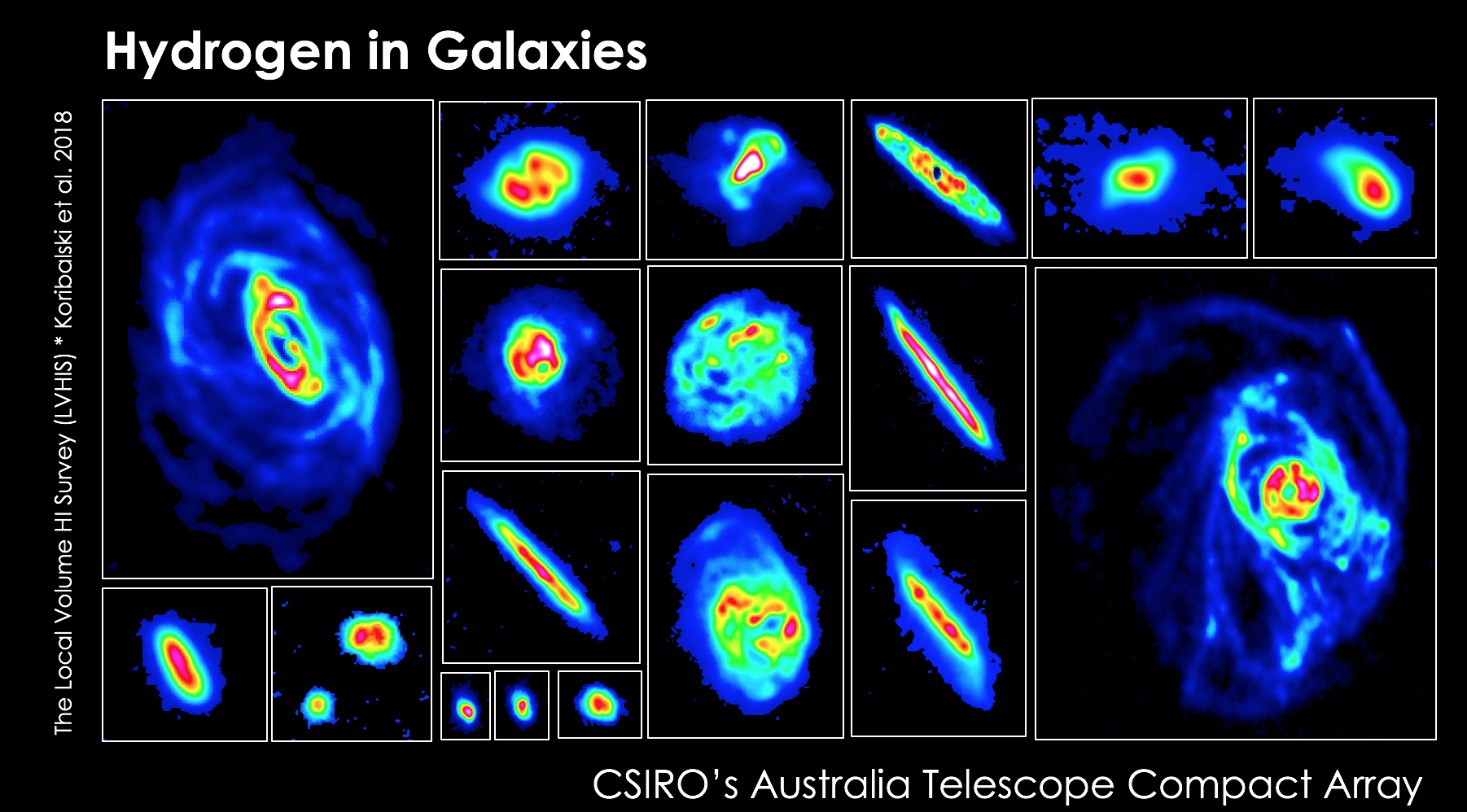}
\caption{LVHIS collage --- ATCA H\,{\sc i} distributions for about a quarter 
   of the LVHIS galaxies (not to scale), showing the diversity of morphologies,
   sizes, and shapes / orientations. The largest LVHIS galaxies are Circinus
   (top left) and M\,83 (bottom right), both spiral galaxies with very 
   extended H\,{\sc i} disks \cite [(Koribalski et al. 2018)]{koribalski18}.}
\label{fig1}
\end{center}
\end{figure}

\begin{figure}[h] 
\begin{center}
 \includegraphics[width=14cm]{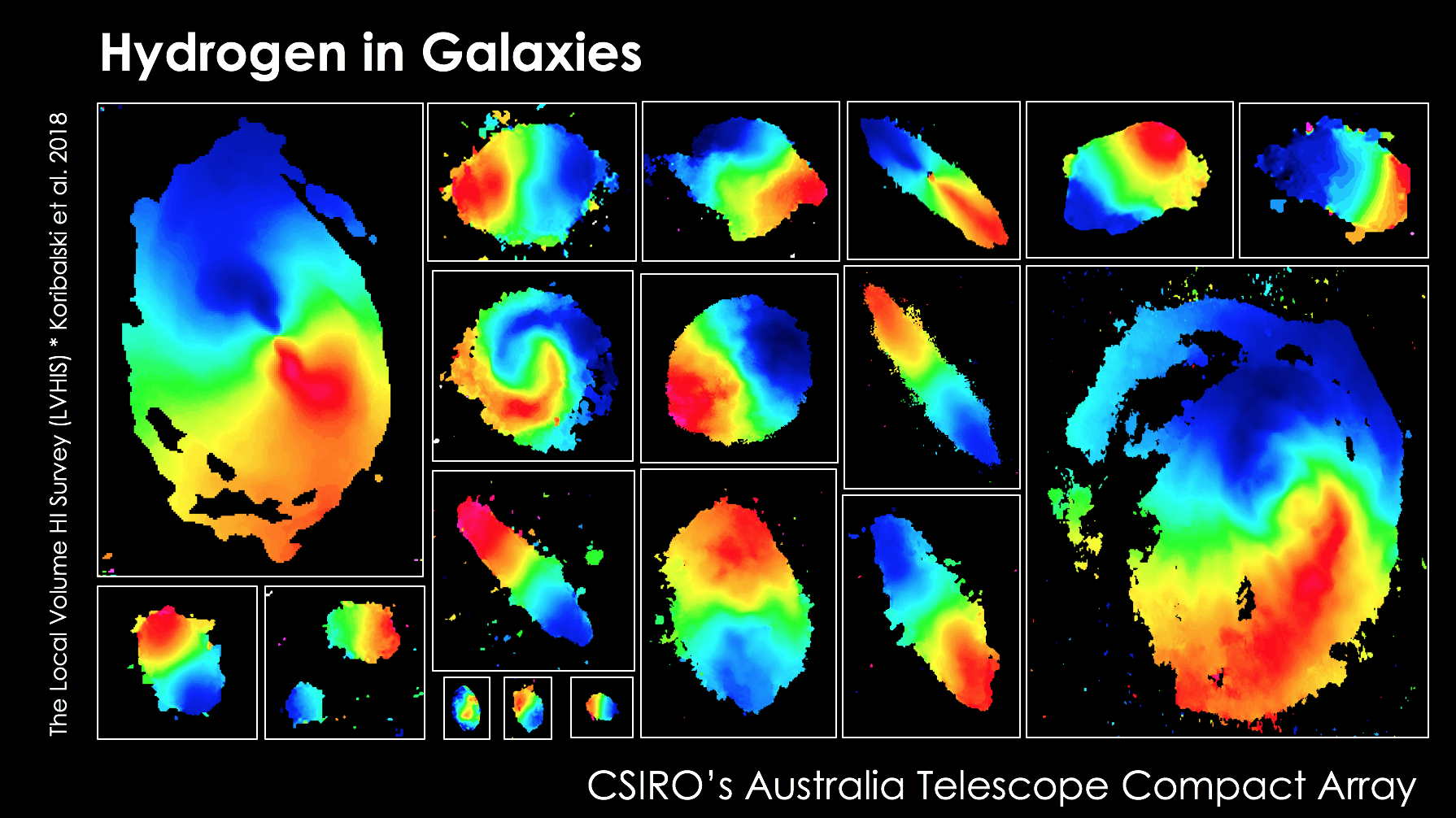}
\caption{LVHIS collage --- ATCA H\,{\sc i} velocity fields, complementing 
   Fig.~1. Most of the LVHIS galaxies show the clear signature of a rotating 
   disk, although peculiar motions and asymmetries are notable. The green 
   colours typically correspond to each galaxy's systemic velocity from which 
   their Hubble distance is derived. The fast rotating disks of spiral galaxies 
   (blue = approaching side, red = receding side) are typically flat in the 
   inner region but can be strongly warped in their outer parts. Using the 
   H\,{\sc i} velocity fields we can determine both the shapes of galaxy disks 
   and their total mass distribution, including their dark matter content as a 
   function of radius. --- For more details see \cite[Koribalski et al. 
   (2018)]{koribalski18} and references therein.}
\label{fig2}
\end{center}
\end{figure}

Fig.~\ref{fig1} shows a collage of LVHIS spiral and dwarf galaxies; their
distributions of cold hydrogen gas typically extend a factor 2 -- 3 beyond 
the bright stellar disks. The red and yellow-coloured reservoirs of dense 
hydrogen gas pinpoint where most of the galaxy's star formation is happening, 
while the dark blue areas indicate large amounts of dormant fuel (cold gas) 
not yet forming stars. We find a veritable zoo of shapes and sizes ranging 
from irregular dwarf galaxies to majestic grand-design spiral galaxies. 
Fig.~\ref{fig2} shows the ATCA H\,{\sc i} velocity fields for the same 
LVHIS galaxies.

\section{WALLABY -- The ASKAP H\,{\sc i} All Sky Survey}
\vspace{0.3cm}

During ASKAP Early Science Phase 1 (Oct 2016 -- Mar 2018) the WALLABY team 
obtained $\sim$800 hours of H\,{\sc i} 21-cm spectral line observations for 
four target fields with an array of up to 16 PAF-equipped antennas. Using 
all 36 beams, typically arranged in a $6 \times 6$ pattern, we achieved the 
desired field of view of 30 square degrees. Our aim was to reach full WALLABY 
depth ($\sim$1.6 mJy\,beam$^{-1}$ per 4~km\,s$^{-1}$ channel), which required 
spending 150-h per field over multiple nights. By targeting nearby, gas-rich 
galaxy groups and clusters we set out to explore the H\,{\sc i} content, 
kinematics and star formation of galaxies as a function of their local 
environment as well as searching for H\,{\sc i} clouds, filaments and plumes 
between galaxies (e.g., \cite[Serra et al. 2015b]{serra15b}, \cite[Saponara 
et al. 2018]{saponara18}). For field number 
one, centered on the NGC~7232 galaxy group, we were able to use only a limited 
ASKAP bandwidth of 48 MHz. Our first WALLABY Early Science results are presented
by Lee-Waddell et al. (2018) on the interacting NGC 7232/3 galaxy triplet and 
surroundings, Kleiner et al. (2018) on the spiral galaxy IC\,5201 and dwarf 
galaxy companions, and Reynolds et al. (2018) on the NGC~7162 galaxy group. 
The second field targeted the Fornax Cluster in Dec 2016; the available 
observing bandwidth was 192 MHz. Our third and fourth fields were centered on 
the Dorado and M\,83 galaxy groups, respectively. WALLABY Early Science results 
are presented by Elagali et al. (2018), who focused on the spiral galaxy 
NGC~1566, with several more papers to follow. ASKAP Early Science Phase 2  
(e.g., mapping H\,{\sc i} line and radio continuum emission, searching for fast
radio bursts) is set to continue with the full ASKAP-36 array and close to 
the full bandwidth of 300 MHz throughout 2019.  \\

WALLABY details can be found at {\em www.atnf.csiro.au/research/WALLABY}, 
including ASKAP updates. We note that the H\,{\sc i} Parkes All Sky Survey 
(HIPASS; Dec $< +25$ degr), which was conducted with an innovative 13-beam 
receiver on the 64-m Parkes Telescope, covers nearly the same sky area as 
WALLABY. While HIPASS has a low resolution of $\sim$15.5~arcmin \& 
18~km\,s$^{-1}$ (\cite[Barnes et al. 2001]{barnes01}) and catalogued more 
than 5000 galaxies (\cite[Koribalski et al. 2004]{koribalski04}, \cite[Meyer 
et al. 2004]{meyer04}, \cite[Wong et al. 2006]{wong2006}), WALLABY --- 
conducted with brand-new, 188-element ASKAP Phased Array Feeds --- has a high
resolution of $\sim$30~arcsec \& 4~km\,s$^{-1}$ and is expected to detect 
over 600\,000 galaxies of which  $\sim$5000 galaxies will be well-resolved
(\cite[Koribalski 2012]{koribalski12}). Interestingly, we already know the 
majority of these well-resolved galaxies (H\,{\sc i} diameter $>$ 150 arcsec) 
quite well as they correspond to the catalogued HIPASS galaxies. We employ 
the remarkably tight H\,{\sc i} size -- mass relation, recently re-visited 
by \cite[Wang et al. (2016)]{wang16}, to estimate the H\,{\sc i} diameters 
of HIPASS galaxies and implications for WALLABY.

\section{Software development: SoFiA, FAT \& 2DBAT}

Our new Source Finding Application (SoFiA; \cite[Serra et al. 2015a]{serra15a}) 
runs on data cubes customized to search for point-like and/or extended 
H\,{\sc i} sources. SoFiA delivers H\,{\sc i} source parameters, spectra, 
moment maps, cublets and masks, and is currently deployed on WALLABY Early 
Science data cubes (e.g., \cite[Lee-Waddell et al. 2018]{lee-waddell18}, 
\cite[Reynolds et al. 2018]{reynolds18}). Most importantly, SoFiA can also 
calculate the reliability of each source candidate. A new function (the 
"busy function"), developed to reliably fit spectral line profiles 
(\cite[Westmeier et al. 2014]{w14}) is also part of SoFiA. 
We also aim to derive detailed kinematical parameters for the well-resolved 
WALLABY galaxies. For this purpose, we are currently testing and comparing 
several algorithms (TiRiFiC, velfit, rotcur, etc.) with the aim to 
develop automated parametrization pipelines. First examples are the Fast 
Automated Tirific (FAT) pipeline developed by \cite[Kamphuis et al. 
(2015)]{kamphuis15} and the 2D Bayesian Automated Tirific (2DBAT) package 
developed by \cite[Oh et al. (2018)]{oh18}, both available on-line.

\end{document}